\documentclass[twocolumn,showpacs,showkeys,floatfix]{revtex4}

\usepackage{graphicx}%
\usepackage{dcolumn}
\usepackage{amsmath}

\newcommand{\xbld}[1]{\mbox{\boldmath $ #1 $}}

\newcommand{\bra}{\langle}
\newcommand{\ket}{\rangle}
\def\NNcha{{{{\rm NN}$^1\!S_0$}}}
\def\SNcha{{{$\Sigma{\rm N}$($T$=3/2)$^3\!S_1$}}}

\begin{document}

\title{Quark Model and Equivalent Local Potential}
\author{Sachiko Takeuchi}
\affiliation{%
Japan College 
of Social Work, Kiyose, Japan}
\author{Kiyotaka Shimizu}
\affiliation{%
Department of Physics, Sophia University, 
Chiyoda-ku, Tokyo, Japan}

\date{\today}%

\begin{abstract}
In this paper, we investigate the short-range repulsion given by the quark 
cluster model employing an inverse scattering problem. We find that the local 
potential which reproduces the same phase shifts to those given by 
the quark cluster model has a strong repulsion at short distances in the 
\NNcha\ channel. 
There, however,  appears an attractive pocket at very short distances due to a 
rather weak repulsive behavior at very high energy region. 
This repulsion-attractive-pocket 
structure becomes more manifest in the channel which has an 
almost forbidden state, \SNcha.  
In order to see what kinds of effects are important 
to reproduce the short-range repulsion in the quark 
cluster model, we investigate the contribution coming 
from the one-gluon-exchange potential and the normalization separately.  
It is clarified that 
the gluon exchange constructs the short-range repulsion in the \NNcha\
while the quark Pauli-blocking effect governs the feature of the repulsive 
behavior in the \SNcha\ channel.
\end{abstract}  

\maketitle

\section{Introduction}
  In nucleon-nucleon (NN) scattering, the phase shift becomes
negative as the relative energy increases.  In order to explain this
behavior, the short-range strong 
repulsion
has been introduced in the NN potential \cite{ham, rei}.
In a microscopic model with the meson-exchange 
potential \cite{mac,nijm},
the vector-meson exchange has been shown to produce 
such a short-range strong 
repulsion.
There also have been many studies 
to investigate the short-range part of the potential
by introducing the subnucleonic degrees of freedom.
Among them, the
model called a quark cluster model (QCM)
\cite{oka, fae, oy, shi, osysup}
is one of the most
successful models which can explain the repulsive
behavior of the phase shift in  baryon-baryon 
scattering.

The characteristic features of the quark model potential are its
nonlocality and energy dependence.  
The former appears by integrating the
internal quark degrees of freedom out
while the latter appears
when interpreting its nonlocality to the energy dependence.
The energy-dependent potential shows that the
core increases as the energy increases \cite{suz}.

By taking parameters to minimize the nucleon mass,
the effect of the orbital [42] symmetry 
to lower the core was found to be diminished \cite{fae}.
This has been confirmed by the calculation which takes into account 
not only $NN$ but also $\Delta \Delta$ and $CC$ channels. 
Therefore it is enough to consider the single channel problem 
to discuss the 
short range part of the baryon-baryon interaction.

It is known that nonlocal 
and energy dependent terms play an 
important role to produce the repulsive behavior of the
NN systems in the conventional
model\cite{par}.
The nonlocality is considered to be important
also in a quark model; it 
has been studied qualitatively 
employing a simple nonlocal potential \cite{shya}.
There, they found that the
approximation
which has usually been used to obtain the energy-dependent local potential
from the nonlocal one is actually valid in the concerning energy 
region.
Furthermore, it was shown that
the equivalent local potential can be obtained from
the simple nonlocal potential
by solving the inverse scattering problem;
namely, they
obtained a unique local potential which gives the same phase shifts as
the nonlocal potential. 

In this paper we employ a realistic baryon potential given by the quark model
 and solve the inverse scattering problem. 
In the $S$-wave NN scattering, 
the norm kernel, which gives rough estimate of the size of the
Pauli-blocking effect,
is known to be small.
The one-gluon-exchange potential 
(OGEP) appearing together 
with the quark exchanges, however,
is large and highly nonlocal in this channel.

On the other hand, there are some channels 
where the norm kernel becomes very 
small. In this case we expect that there appears a large repulsive interaction due 
to the Pauli blocking effect \cite{osy}. 
In order to look into
these two effects in details, we study two typical channels, 
\NNcha\  and \SNcha,
and investigate each 
contribution coming from OGEP and the norm kernel separately.

In the next section we explain briefly the quark cluster model.
The method to obtain the baryon potential from QCM (the QCM potential)
is 
also discussed.
In section 3,
we explain the inverse scattering problem;
we use this method to derive
the energy-independent local potential which reproduces the 
same phase shifts as those obtained
from the QCM potential.
We explain two types of the quark cluster models in section 4. 
One is called the OGEP quark model, where 
OGEP plays the dominant role to reproduce the mass difference between nucleon 
and $\Delta$, and also to produce the NN short-range repulsion. 
The other is the 
hybrid chiral (HC) quark model where the pseudoscalar- and 
scalar-meson-exchange 
potential between quarks including quark exchanges 
are taken into account together with OGEP.  
Numerical results are given in section 5.
It is shown that 
the obtained local potential has a strong short-range repulsion.
It, however, also has an attractive pocket at very short distances 
which reflects that the nonlocal repulsion becomes weaker 
as the energy increases.
It is also shown that the channel which has an 
almost forbidden state has such a structure in more
extensive way.
Also, the nonlocality seems to become more important there.
Summary is given in section 6.

\section{Quark Cluster Model}
Here we briefly summarize the quark cluster model (QCM) to study the baryon-baryon 
scattering in terms of the constituent quarks. 

The total wave 
function of the six quark system is given by
\begin{equation}
\Psi(\xbld{\xi}_A ,\xbld{\xi}_B, 
{ \xbld{R}}_{AB})={\cal A}[\phi_A(\xbld{\xi}_A)
\phi_B(\xbld{\xi}_B)\chi({\xbld{R}}_{AB})],
\end{equation}
where $\phi_A$ and $\phi_B$ are the single baryon wave function for baryons 
$A$ and $B$. They are given by a product of orbital, 
flavor-spin and color parts as
\begin{equation}
\phi(\xbld{\xi})=\varphi(\xbld{\xi})S([3]_{f \sigma})C([111]_c).
\end{equation}
The $\xbld{\xi}_A$ and $\xbld{\xi}_B$ are internal coordinates of the 
baryon $A$ and $B$,  $\chi$ is the relative wave function, 
$\xbld{R}_{AB}$ is the relative coordinate between the baryons 
$A$ and $B$ and ${\cal A}$ 
is the antisymmetrization operator among six quarks.
Assuming that the internal wave function $\phi(\xbld{\xi})$ is known, 
we obtain the following RGM equation 
to determine the relative wave function 
$\chi$,
\begin{equation}
\int H(\xbld{R},\xbld{R}')\chi(\xbld{R}')d\xbld{R}'
= E \int N(\xbld{R},\xbld{R}')\chi(\xbld{R}')d\xbld{R}',
\label{RGMeq}
\end{equation}
where $H$ and $N$ are the Hamiltonian and  normalization kernels.
\begin{widetext} 
\begin{eqnarray}
\left\{ \begin{array}{c}
H(\xbld{R},\xbld{R}')\\
N(\xbld{R},\xbld{R}')
\end{array} \right\}
= \int d\xbld{\xi}_Ad\xbld{\xi}_Bd\xbld{R}_{AB} 
\phi^\dagger_A(\xbld{\xi}_A)\phi^\dagger_B(\xbld{\xi}_B)
\delta(\xbld{R}-\xbld{R}_{AB})
\left\{ \begin{array}{c}
H\\ 1
\end{array} \right\} 
{\cal A} [\phi_A(\xbld{\xi}_A)\phi_B(\xbld{\xi}_B)
\delta(\xbld{R}'-\xbld{R}_{AB})]~.
\end{eqnarray}
\end{widetext}
Employing the following gaussian,
\begin{equation}
g(\xbld{r},b)=(\sqrt{\pi}b)^{-3/2} \exp \{-r^2/(2b^2) \},
\end{equation}
we take the orbital part of the internal wave function  
$\varphi(\xbld{\xi})$ as
\begin{equation}
\varphi(\xbld{\xi})=g(\xbld{\xi}_1,\sqrt{2}b)g(\xbld{\xi}_2,\sqrt{3/2}b),
\end{equation}
where
\begin{equation}
\xbld{\xi}=(\xbld{\xi}_1,\xbld{\xi}_2)= \left \{ 
\begin{array}{c}
(\xbld{r}_1-\xbld{r}_2, \frac{\xbld{r}_1+\xbld{r}_2}{2}-\xbld{r}_3) 
\hspace{5pt} \mbox{ for }A \\
(\xbld{r}_4-\xbld{r}_5, \frac{\xbld{r}_4+\xbld{r}_5}{2}-\xbld{r}_6) 
\hspace{5pt} \mbox{ for }B \\
\end{array} \right..
\end{equation}
In this case, the norm kernel is given by the following equation.
\begin{widetext} 
\begin{eqnarray}
N(\xbld{R},\xbld{R}')&=&\delta(\xbld{R}-\xbld{R}') 
-9\bra P_{36}^{(f \sigma c)}\ket
\left({27 \over 16\pi b^2}\right)^{3\over2}
\exp\left[-{15\over 16 b^{2}}(R^{2}+R'{}^{2})+{9\over 8 b^{2}}
\xbld{R}\cdot\xbld{R}'\right]\\
&=&\sum_{n,\ell=0}^\infty  [1-9\bra P_{36}^{(f \sigma c)}\ket
(\frac{1}{3})^{2n+\ell}] \sum_{m=-\ell}^\ell 
u_{n\ell m}(\xbld{R})u_{n\ell m}(\xbld{R}')^{\ast},
\label{eq:HOexpansion}
\end{eqnarray}
\end{widetext} 
where $u_{n\ell m}(\xbld{R})$ is the $n\ell m$ 
harmonic oscillator wave function with the size 
parameter $\sqrt{2/3}b$, and $P_{36}^{(f \sigma c)}$ is the permutation 
operator of the 3rd and 6th 
quarks in the flavor, spin and color spaces.
The expectation values of the permutation operator $P_{36}^{(f \sigma c)}$ 
are recited in table \ref{tab0} for each of the 
\NNcha\  and the \SNcha\ channels.
\begin{table}
\caption{\label{tab0}Coefficients of $n=0~\ell=0$ state}
\begin{ruledtabular}
\begin{tabular}{ccc} 
BB  & $\bra P_{36}^{(f \sigma c)}\ket$ & 
$1-9\bra P_{36}^{(f \sigma c)}\ket$ \\ \hline
\NNcha & $-\frac{1}{81}$ & $\frac{10}{9}$ \\ 
\SNcha & $\frac{7}{81}$ & $\frac{2}{9}$ \\ 
\end{tabular}
\end{ruledtabular}
\end{table}

To see the rough size of the quark effects on the spin-flavor-independent
observables, 
it is useful to see the matrix element of the 
exchange operator in the flavor-spin-color space.
This corresponds to the normalization of the
relative $0s$ state, which is affected most largely
by the internal degrees of freedom.
As seen from the table \ref{tab0}, 
we expect that the effect of Pauli blocking 
is not important for the \NNcha\ channel, because the factor 
$(1-9\bra P_{36}^{(f \sigma c)}\ket)$ is 10/9, which is close to 1.
On the other hand, the factor is 2/9, 
much smaller than 1, in the \SNcha\ channel. 
Suppose this factor is found to be zero, it indicates 
that there is a forbidden state in the concerning channel. 
Thus, 
the $0s$ state in the \SNcha\ channel can be called as an almost forbidden 
state. We will later discuss the role of this almost forbidden 
state on the phase shift and the equivalent local potential.

Here we present two ways to rewrite the RGM equation (\ref{RGMeq}) into 
``the Schr\"{o}dinger equation."
One is to put the exchange part of the norm kernel into the hamiltonian,
the other is to divide the equation by the norm kernel.
We explain them in the following.

The norm kernel and hamiltonian kernel can be decomposed as
\begin{eqnarray}
N&=&1+N_{ex}\\
H&=&K_d+K_{ex} +V~.
\end{eqnarray}
where $N_{ex}$ is the exchange part of the norm kernel,
$K_{d}$ and $K_{ex}$ are direct and  
exchange parts of the kinetic energy, respectively. 
The potential $V$ is a part 
due to the quark-quark potential term together with the quark exchanges.
The RGM equation (\ref{RGMeq}) can be written in the Schr\"{o}dinger equation: viz.,
\begin{equation}
H_s\chi=E\chi, \hspace{20pt} H_s=H-EN_{ex}. 
\label{eq:EdepH}
\end{equation}
Since the $K_{ex}$ and $V$ are highly nonlocal due to the 
quark exchanges,
the hamiltonian $H_s$ becomes nonlocal as well as energy dependent.

In order to avoid the energy dependent hamiltonian, 
we can rewrite the RGM 
equation in a different way:
\begin{equation}
H \chi= EN \chi \rightarrow 
\frac{1}{\sqrt{N}}H\frac{1}{\sqrt{N}} \sqrt{N}\chi =\sqrt{N} \chi.
\end{equation}
Then the Schr\"odinger equation becomes
\begin{equation}
\tilde H\psi = E\psi.
\label{qmp}
\end{equation}
Here the energy independent hamiltonian $\tilde H$ is defined
in the following way:
\begin{eqnarray}
\tilde H &=& \frac{1}{\sqrt{N}}H\frac{1}{\sqrt{N}}\\
 &=&K_d + \tilde V_{QCM},
\end{eqnarray}
where
\begin{eqnarray}
\tilde V_{QCM} &=& 
\left(\frac{1}{\sqrt{N}}(K_d+K_{ex})\frac{1}{\sqrt{N}}-K_d\right)
+ 
\frac{1}{\sqrt{N}}V\frac{1}{\sqrt{N}}. \nonumber\\
\end{eqnarray}
$\tilde V_{QCM}$  can be considered 
as the potential term 
in the usual Schr\"{o}dinger equation for the baryon-baryon scattering,
which is very nonlocal, but not energy dependent.
The RGM wave function can be obtained from $\psi$ as $\chi=N^{-1/2}\psi$.

The equations (\ref{RGMeq}),  (\ref{eq:EdepH}), and (\ref{qmp}) are equivalent to among each other
provided that $N^{-1/2}$ is well-defined. Their phase shifts are the same
and give the same equivalent local potential, 
which we will discuss in the next section.
Though the former treatment is more intuitive, 
the obtained potential depends strongly
on the energy \cite{Ta02}.  
When  looking into the nonlocality of the potential,
the latter treatment has an advantage that it does not depend on the energy.

The origin of the short-range repulsion has been argued to come from the 
quark potential and/or the quark Pauli principle; 
which of these two reasons
is more important depends on the channel.  
The effect of the nonlocal potential has been 
discussed employing the simplified model in ref.\ \cite{shya}. 
As for the effect 
of the Pauli principle, we expect that the normalization
 $\bra N\ket$ gives a 
rough estimate. 
In the channel where $\bra P_{36}^{(f \sigma c)}\ket$ is positive,
 $\bra N^{ex}\ket$ is negative,
 which causes repulsion in
the short-range part of the potential $\tilde V_{QCM}$. 
This can be understood like this:
negative $\bra N_{ex}\ket$ means that the short-range part 
are partially forbidden by the quark Pauli principle. 
When the wave function is expanded by the 
harmonic oscillator as in eq.\ (\ref{eq:HOexpansion}), 
this effect appears 
mainly in the $n$=0 component, because the effect is reduced by 
a factor of $(\frac{1}{9})^{n}$.  The effect appears not only in the 
normalization but also in the kinetic energy part.  Then the diagonal 
part of the effective kinetic energy 
$\frac{1}{\sqrt{N}}(K_d+K_{ex})\frac{1}{\sqrt{N}}$ 
remains the same as that of 
$K_d$, but the non-diagonal part becomes weaker.  
As a result the 
mixing between the $0\ell$ and $1\ell$ components becomes smaller and so 
 the energy for the $0\ell$ state  
becomes higher than the one in the case of 
 $N_{ex}=0$ \cite{TS02}. See appendix B.

Since the  $\bra N^{ex}\ket$ is positive in the \NNcha\ channel, 
we expect that the effect of the 
Pauli principle does not produce the short-range repulsion there. 
There exist some 
channels which have a large negative $\bra N^{ex}\ket$, 
such as the \SNcha\ channel as shown in the 
table.

\section{Energy independent local potential}
In this section, we apply the inverse scattering method 
\cite{Newton, AM63} 
to obtain the energy independent local 
potential which gives the same phase shifts given by the quark 
cluster model. 
This equivalent local potential shows us  more intuitive picture 
of the nature of this nonlocal potential.
Since we are mainly interested in the short-range behavior of the 
potential, we look only into the $S$-wave scattering here. 

All the information 
on the scattering observables is in the S-matrix $S(k)$.
Once $S(k)$ is known, the potential 
$V(r)$ is obtained by the following equation called 
Marchenko equation \cite{Newton, AM63}.
First we calculate the following $F(r)$ 
from the S-matrix with poles at \{$k=i\kappa_j$\},
\begin{equation}
F(r)=-\frac{1}{2\pi} \int_{-\infty}^{+\infty} e^{ikr} \{ S(k)-1 \} dk
+\sum_{{\rm all}~\kappa_j>0} c_j^2e^{-\kappa_j r},
\label{eq22}
\end{equation}
where $c_j^2$ is
\begin{equation}
c_j^2=\mbox{Residue} \{ S(k) \}
\mbox{ at } k=i\kappa_j~(\kappa_j>0).
\end{equation}
Next we solve the following integral equation using the $F(r)$.
\begin{equation}
K(r,r') = -F(r+r')-\int_r^{\infty} F(r+r'')K(r,r'') d r''. 
\label{eq24}
\end{equation}
Then the potential $V(r)$ is given by
\begin{equation}
2\mu V(r) = -2 \frac{d}{dr}K(r,r),
\label{eq25}
\end{equation}
where $\mu$ is the reduced mass for the baryon-baryon scattering.

Provided that the potential is local,
the potential $V$ in eq.\ (\ref{eq25}) 
can be constructed uniquely from the given S-matrix.
Also, for any local potentials $V_1$, $V_2$ and $V=V_1+V_2$, 
suppose we obtain S-matrix for each potential 
and reconstruct $V'_1$, $V'_2$ and $V'$
by this method, 
then $V'=V'_1+V'_2$ holds because $V=V'$ {\it etc}.\ hold.
This is not the case if one of $V_i$ is nonlocal.
We call this the potentials are ``additive'', and 
use the deviation $V'-(V'_1+V'_2)$ 
as the rough estimate of the degrees of the nonlocality
later in section \ref{sec:5.3}.

This method has been applied on the baryon-baryon scattering employing the 
 simple gaussian-type nonlocal potential \cite{shya}.
In this paper, we employ this method to the QCM potential, which
will be defined in the next section.
The procedure is (1) to obtain the S-matrix $S(k)$ from the QCM potential
up to very high momentum ($k \sim 15$ fm$^{-1}$), (2) to construct 
a local potential from the $S$-matrix by using eqs.\ (\ref{eq22}-\ref{eq25}).
See appendix A for the detail.

\section{Quark model Hamiltonian}
The Hamiltonian of the nonrelativistic quark model \cite{close} 
is the sum of the kinetic energy 
and two body interaction:
\begin{equation}
H^q=\sum_i \left(m_i+\frac{\xbld{p}_i^2}{2m_{i}}\right)-K_{cm} + V^q,
\end{equation}
where $K_{cm}$ is the kinetic energy of the center of mass motion. 
The two-body 
potential $V^q$ may consist of the pseudoscalar- 
and scalar-meson-exchange potentials 
as well as of the confinement and one-gluon exchange (OGEP).
These potentials have been employed to describe the single baryon structure
\cite{isgur}.
The color-magnetic part of OGEP \cite{ruju} is known to reproduce
the mass difference between octet and decuplet baryons.
On the other hand, the chiral quark model \cite{glozm1}
includes the pseudoscalar meson-exchange potentials,
which also contribute to the mass difference.
The pseudo-scalar mesons appear
as the Goldstone bosons together with their chiral partner $\sigma$ meson.

In the following,  
we employ two types of the quark models. One is the OGEP quark model
for the baryon-baryon scattering, 
where only the long-range part of the meson-exchange potentials are 
included in addition to the OGEP. 
The other one is the hybrid chiral (HC) quark model, where 
the pseudoscalar- and scalar-meson exchanges occur between quarks.
Thus the potential $V^q$ becomes
\begin{equation}
V^q=\sum _{i>j} V_{ij}=V^{conf}_{ij}+V^{OGEP}_{ij}
+V_{ij}^{\sigma}+V_{ij}^{ps}.
\end{equation}
The explicit form for each potential is as follows.
We take the two-body confinement term:
\begin{equation}
V^{conf}_{ij}(\xbld{r}_{ij})=
-\xbld{\lambda}_i \cdot \xbld{\lambda}_j a_cr^2_{ij}, 
\end{equation}
where $\xbld{\lambda}_i$ is the color SU(3) generator of the $i$-th quark.
OGEP consists of the color coulomb, electric and magnetic terms:
\begin{eqnarray}
\lefteqn{V^{OGEP}_{ij}(\xbld{r}_{ij})=}
\nonumber\\&&
\xbld{\lambda}_i \cdot \xbld{\lambda}_j 
\frac{\alpha_s}{4}
\left \{ \frac{1}{r_{ij}}-\xi_{ij}\frac{\pi}{m^2}
(1+\frac{2}{3}\xbld{\sigma}_i\cdot\xbld{\sigma}_j)\delta(\xbld{r}_{ij})
\right \}~.
\end{eqnarray}
Here we have introduced the flavor SU(3) breaking factor $\xi_{ij}$. 
This factor takes the following value:
\begin{equation}
\xi_{ij}=
\left \{ \begin{array} {cl}
1 & \mbox{for } i, j \neq s \hspace{5pt}\mbox{quark}\\
\xi_1 & \mbox{for } i \mbox{ or } j = s \hspace{5pt}\mbox{quark}\\
\xi_2 & \mbox{for } i \mbox{ and } j = s \hspace{5pt}\mbox{quark}\\
\end{array} \right..
\end{equation}
The parameters $\xi_1$ and $\xi_2$ are fixed so as to reproduce the empirical
octet baryon masses.
The following scalar- and 
pseudoscalar-meson-exchange potentials are taken into account. 
\begin{eqnarray}
V_{ij}^{\sigma}(\xbld{r}_{ij}) &=& 
-\frac{g_\sigma^2}{4 \pi} \frac{\Lambda^2}{\Lambda^2-m_{\sigma}^2}
\left( \frac{e^{-m_{\sigma}r_{ij}}}{r_{ij}}
-\frac{e^{-\Lambda r_{ij}}}{r_{ij}} \right), \\
 V_{ij}^{ps}(\xbld{r}_{ij}) &=& \frac{1}{3} 
\frac{g_c^2}{4 \pi} \frac{m_{ps}^2}{4m^2} 
\frac{\Lambda^2}{\Lambda^2-m_{ps}^2}
\xbld{f}_i \cdot \xbld{f}_j \xbld{\sigma}_i\cdot\xbld{\sigma}_j
\nonumber\\&\times&
\left \{ \frac{e^{-m_{ps}r_{ij}}}{r_{ij}}
-\left( \frac{\Lambda}{m_{ps}} \right)^2
\frac{e^{-\Lambda r_{ij}}}{r_{ij}} \right \}.
\end{eqnarray}
We introduce the cut-off $\Lambda$ for the meson-exchange potentials.
The SU(3)-octet pseudoscalar mesons, $\pi$, K and $\eta$, are included. 

Both of the models we employ include a few parameters.
In the OGEP quark model, the quark 
mass $m$ and size parameter $b$ are taken to be 313 MeV and 0.6 fm.
The quark-gluon coupling constant $\alpha_s$ and confinement strength $a_c$ 
are fixed by the nucleon and $\Delta$ mass difference and the stability 
condition for the nucleon against the variation of the size parameter $b$
\cite{fae}. 
The long-range parts of the 
meson-exchange potentials are included as an interaction between baryons
added to the potential term in eq.\ (\ref{qmp}) in order to 
reproduce the phase shifts at low energies. 
The quark-$\sigma$-meson coupling constant is adjusted to reproduce the  
\NNcha\ scattering phase shift at the peak. 
These parameters are given in table \ref{tab1}. 
%
\begin{table}
\caption{\label{tab1}Parameters of OGEP Quark Model (OGEP) and
Hybrid Chiral Quark Model (HC)}
\begin{ruledtabular}
\begin{tabular}{ccccccccccc} 
\multicolumn{5}{l}{OGEP and HC}\\
\multicolumn{6}{c}{mass (in MeV)}&\multicolumn{2}{c}{scale (in 
fm$^{-1}$)}\\
m & $m_{\sigma}$ & $m_{\pi}$& $m_K$ & $m_{\eta}$ &&$b$ &$\Lambda$\\
313 &675 & 139 & 494 & 547 && 0.6& 4.2 \\
\hline
& $\alpha_s$ & $\xi_1$ & $\xi_2$ & \multicolumn{2}{c}{$a_c$ [MeV/fm$^2$]}
&$\frac{g_c^2}{4 \pi}$ & $\frac{g_\sigma^2}{4 \pi}$\\
OGEP & 1.517 & 0.603 & 0.110 &\multicolumn{2}{c}{26.6} & 0.592 & 0.782 \\ 
HC~~~ &  1.003 & 0.683 & 0.258 & \multicolumn{2}{c}{11.6} & 0.592 & 
0.956 \\ 
\end{tabular}
\end{ruledtabular}
\end{table}
In the table 
the coupling constants of 
the meson-exchange potentials are given in terms of the meson-quark coupling;
 the baryon-meson 
 coupling constant can 
be calculated from the meson-quark coupling constant and the quark 
 distribution in the baryon.

In the HC quark model, 
the quark-gluon coupling constant $\alpha_s$ is fixed to 
reproduce the nucleon and $\Delta$ mass difference together with the 
pseudoscalar-meson-exchange potentials \cite{glozm1, shiglo}.
 Note that the coupling constant 
$\alpha_s$ becomes smaller than the one in the OGEP quark model because the 
pseudoscalar-meson exchange also contributes to the mass difference.
The quark-exchange terms for the 
meson-exchange potentials are also included. 
The parameters used in the present model are also given 
in table \ref{tab1}.

\section{Results}
\subsection{Phase shifts and equivalent local potential}
In this section, we show the numerical results of the scattering phase shifts 
and their equivalent local potentials. 

\begin{figure}[bp]
\includegraphics[width=7cm,clip]{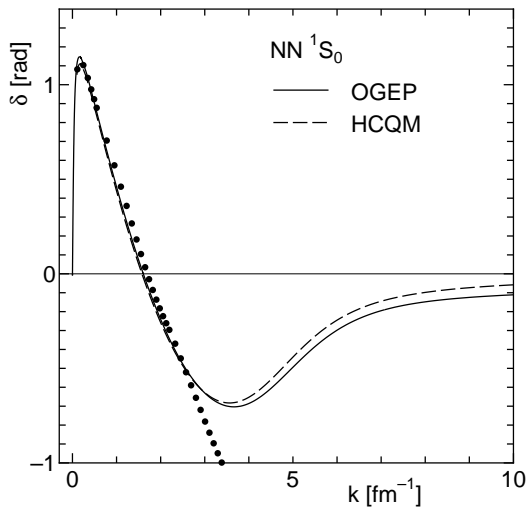}
\caption{Phase shifts for the \NNcha\ channel. 
Dots are experimental values.
OGEP and HCQM stand for the OGEP quark model, where only gluon 
exchange is considered between quarks, and the HC quark model, where
the pseudoscalar and scalar meson exchanges 
are also included between quarks,
respectively.}
\label{fig1}
\end{figure}

First we 
show the results of the phase shift for the \NNcha\ channel 
in Fig.\ \ref{fig1} 
together with the observed phase shift.
%
We employ the OGEP and HC 
quark models to obtain the QCM phase shifts. 
Both models reproduce the phase shifts up to the momentum $k$=2.5 fm$^{-1}$.
Both of the phase shifts, however, go to zero rapidly 
when the $k$ becomes beyond about 5 fm$^{-1}$. 
As will be seen later, 
this weak repulsive behavior at high energies partially originated
from the nonlocality of the potential.

\begin{figure}
\includegraphics[width=7cm,clip]{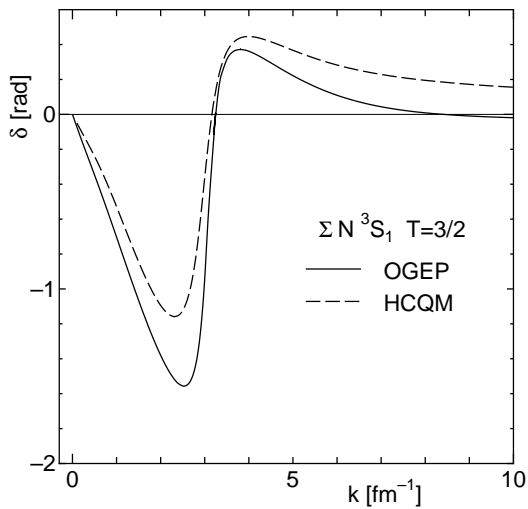}
\caption{Phase shifts for the \SNcha\ channel. 
For further explanations see Fig.\  \ref{fig1}.}
\label{fig2}
\end{figure}

In Fig. \ref{fig2}, the phase shift for \SNcha\
channel is shown.  
%
From the figure, we see that there exists a strong repulsion even at low 
energies. This is due to a smallness of the norm kernel in this channel.
Though the original quark potential is different, 
the OGEP quark model and HC quark model give results similar to each other.

\begin{figure}
\includegraphics[width=7cm,clip]{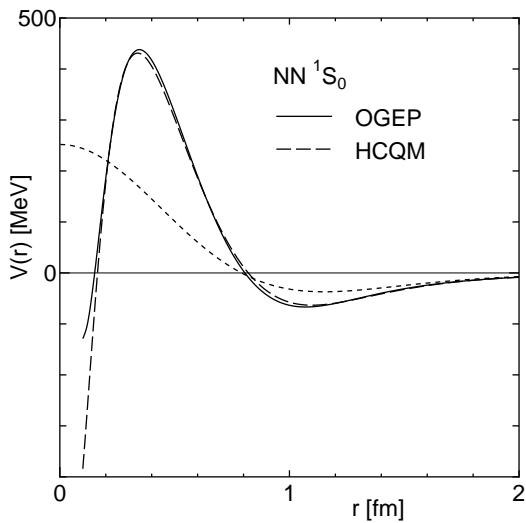}
\caption{Equivalent local potential for the \NNcha\ channel using the OGEP 
or HCQM models. 
Dotted line indicates the local part of the quark model potential 
in the OGEP model.}
\label{fig3}
\end{figure}

\begin{figure}
\includegraphics[width=7cm,clip]{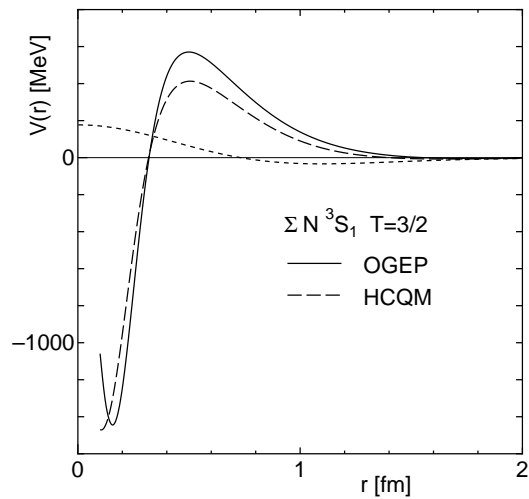}
\caption{Equivalent local potential for the \SNcha\ channel. 
For further explanations see Fig.\ \ref{fig3}.}
\label{fig4}
\end{figure}

In Figs.\ \ref{fig3} and \ref{fig4},
we show the results of the equivalent local potential given by 
solving the inverse scattering problem.
%
%
There,
dotted lines denote the local part of the original 
potential of the OGEP quark model. 
We see that the nonlocal part of the quark model potential 
plays an important role especially at short distances. 
It is noteworthy that there
are very-short-range attraction in addition to the
usual short-range repulsion in both of the channels.
This reflects that the nonlocal part, which is repulsive at the intermediate region, 
reduces effectively at the very high energy region.
In the NN system, some of the conventional soft-core models
give a similar attraction \cite{sriva}.
This property is more manifest in the \SNcha\ channel.
There, the very-short-range attraction 
is due to an existence of 
the almost forbidden state, 
which will be discussed later in more details.   

\subsection{Roles of norm kernel and OGEP}

In order to see what kinds of effects 
are important to produce the short-range 
repulsive interaction in the quark cluster model, we investigate 
 contributions from norm kernel and OGEP separately in the OGEP model.
The potential term in the OGEP quark model
in eq.(\ref{qmp}) can be rewritten as
\begin{eqnarray}
V &=& \tilde V_{QCM} + V_{m}\\
\tilde V_{QCM} &=& \tilde V_{K/N}+\tilde V_{G/N} \\
\tilde V_{K/N} &=&
\frac{1}{\sqrt{N}}(K_d+K_{ex})\frac{1}{\sqrt{N}}-K_d\\
\tilde V_{G/N} &=&\frac{1}{\sqrt{N}}V_{OGEP}\frac{1}{\sqrt{N}}
\end{eqnarray}
We denote the above three terms  
as $K/N$ for $\tilde V_{K/N}$, $G/N$ for $\tilde V_{G/N}$ and $V_m$
in the figures.
Note that the meson-exchange term is taken into account 
as the baryon-baryon interaction in the OGEP quark model.

\begin{figure}
\includegraphics[width=7cm,clip]{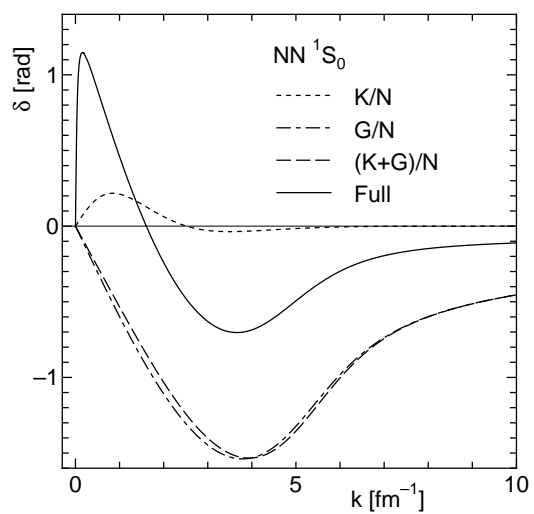}
\caption{Calculated phase shifts for the \NNcha\ channel 
in the OGEP quark model.
$K/N$ denotes the contribution from the normalization and 
kinetic energy terms,
$G/N$ denotes the contribution from the normalization and OGEP terms, and
$(K+G)/N$ is the contribution from the normalization, kinetic energy 
and OGEP terms. Full is the contribution from $(K+G)/N$ and meson-exchange 
potential $V_m$.}
\label{fig5}
\end{figure}

We first show the calculated phase shift from this potential 
for the \NNcha\ channel in
Fig.\ \ref{fig5}. 
%
Here each contribution from $K/N$ or $G/N$ is shown. 
We see 
that the contribution from the norm kernel $K/N$ 
is rather weak and attractive at 
low energies.  This is because of a small enhancement of 
the norm kernel due to the quark exchange as seen in table \ref{tab0} 
\cite{oy, shi, TS02}.
The contribution from the OGEP, namely, $G/N$ is strongly repulsive, and
the contribution from both terms $(K+G)/N$ is almost the same as $G/N$. 
Therefore we can conclude that the most important term to reproduce the 
short-range repulsion in the \NNcha\ channel is OGEP. 
When the meson-exchange potential, $V_{m}$, is taken into account, 
the phase shift becomes positive at low energies as shown by the solid line 
(denoted as Full in the figure).

\begin{figure}
\includegraphics[width=7cm,clip]{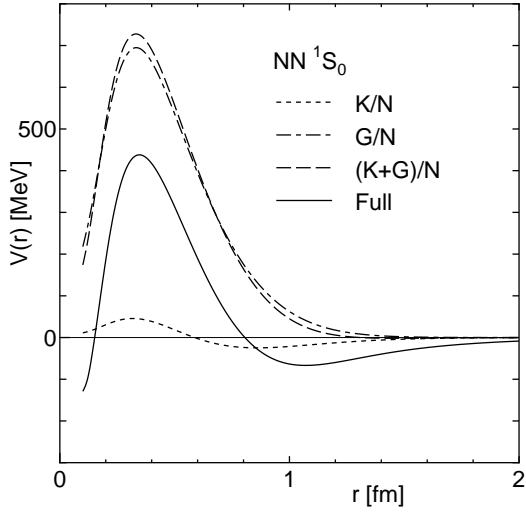}
\caption{Equivalent local potential for the \NNcha\ channel in 
OGEP quark model. 
For further explanations see Fig.\ref{fig5}.}
\label{fig6}
\end{figure}

Employing the calculated phase shifts shown 
in Fig.\ \ref{fig5}, the equivalent 
local potential is calculated for \NNcha\ 
by solving the inverse scattering problem.
The results are shown in Fig.\ \ref{fig6}. 
%
As seen in the figure, there exists 
a strong repulsion due to the norm and OGEP terms, 
$G/N$, at short distances.
It, however, becomes weak at very short distances;
 there appears an attractive pocket at short distances.
 By including the meson exchange 
potential consisting of $\sigma$, $\pi$ 
and $\eta$ meson-exchange potentials, the 
long-range part of the potential becomes attractive.

\begin{figure}
\includegraphics[width=7cm,clip]{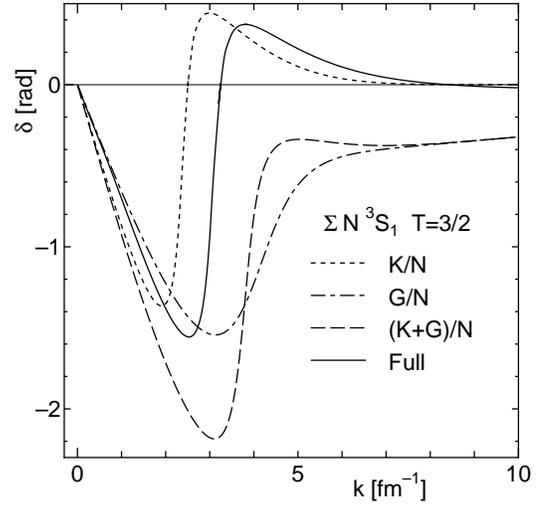}
\caption{Calculated phase shifts for the \SNcha\ channel in 
OGEP quark model. 
For further explanations see Fig.\ref{fig5}.}
\label{fig7}
\end{figure}

\begin{figure}
\includegraphics[width=7cm,clip]{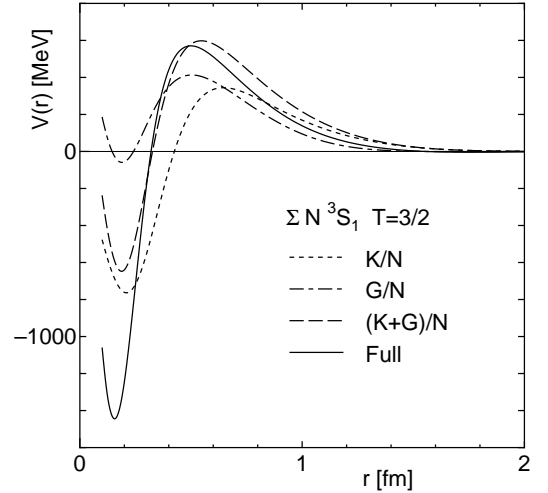}
\caption{Equivalent local potential for the \SNcha\ 
channel in the OGEP quark model. 
For further explanations see Fig.\ref{fig5}.}
\label{fig8}
\end{figure}

The results for the \SNcha\ channel are shown in Figs.\ 
\ref{fig7} and \ref{fig8} by employing the OGEP quark model. 
%
%
The reason why 
we are interested in this channel is that there exists the almost forbidden 
state; we can study the role of the Pauli blocking effect more 
clearly in this channel. In Fig.\ \ref{fig7}, 
we show the contributions from 
$K/N$, $G/N$, $(K+G)/N$ and 
Full calculation which includes the meson-exchange 
potential $V_m$.  
As seen in the figure, $K/N$ 
gives the negative phase shift at the low energy region, 
which indicates strong repulsion
at long ranges.
It is also interesting to see the phase shifts increase sharply around 
$k\sim 2 - 4$ fm$^{-1}$. 
These are due to the Pauli-blocking effect, 
because this sharp increase of the phase shift is seen in the cases 
which include the $K/N$. 
Since
$\bra P_{36}^{(f \sigma c)}\ket$ is $\frac{7}{81}$ in this channel,
the $0s$ component of the norm kernel is
 reduced 
by a factor of $1-7/9=2/9$. 
Similarly, the $0s$-$1s$ off-diagonal component of the kinetic term 
and the $0s$ diagonal component are reduced
by this factor 2/9.
After being divided by the norm kernel, diagonal part of 
the kinetic energy term 
$K/N$ are not reduced, but the $0s$-$ns$ non-diagonal parts are reduced 
by a factor of $\sqrt{2/9}$. 
This smaller-than-1 factor causes less mixing
between the $0s$-$1s$ component.
Thus, the $K/N$ term has a node in the phase shift
with the repulsion at the lower energy region\cite{TS02}.
See appendix B, where 
this mechanism is imitated by a simple model.
Suppose the factor is zero, indicating the system has a forbidden 
state, the phase shift decreases continuously toward $-\pi$ at $k= 
\infty$ in order to satisfy the Levinson's theorem.

This feature is also seen in the equivalent local potential shown in 
Fig.\ \ref{fig8}.
In the equivalent local potential, 
there appears a very strong attractive 
potential at short distances with a strong repulsion in the 
intermediate range. 
There is a quasi-bound state in this attractive pocket, which 
corresponds to the sharp increase of the phase shift.
If the factor were zero, the resonance would become
a bound state; which is the way to describe a forbidden state in 
terms of a local potential because all other real states
are forced to be orthogonal to the bound state.

\subsection{Roles of nonlocality}
\label{sec:5.3}

\begin{figure}
\includegraphics[width=7cm,clip]{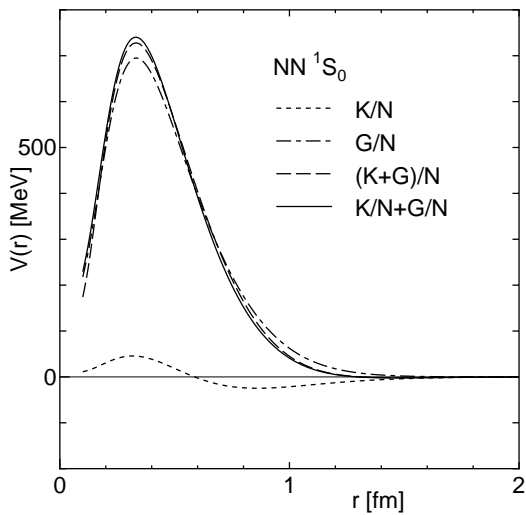}
\caption{Equivalent local potential for the \NNcha\ channel in 
OGEP quark model.
$(K+G)/N$ and $K/N+G/N$ are shown for a comparison. 
For further explanations see Fig.\ref{fig5}.}
\label{fig9}
\end{figure}

\begin{figure}
\includegraphics[width=7cm,clip]{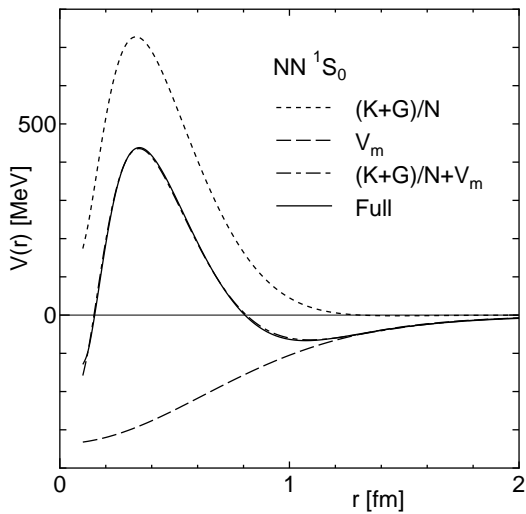}
\caption{Equivalent local potential for the \NNcha\ channel in 
OGEP quark model.
$(K+G)/N$ + $V_m$ and Full are shown for a comparison. 
For further explanations see Fig.\ref{fig5}.}
\label{fig10}
\end{figure}

In order to estimate the degrees of the 
nonlocality, 
we investigate whether the contributions from $K/N$, $G/N$ and $V_m$ 
are additive or not.  Each local potential given by solving the 
inverse scattering and their sum are plotted in Figs.\ 
\ref{fig9} and \ref{fig10}.
%
%
In Fig.\ \ref{fig9}, we see that 
the sum of local potentials $V'_{K/N}+V'_{G/N}$ (solid line) is 
similar to $V'_{(K+G)/N}$ 
(dashed line), so the nonlocality is 
weak as we mentioned in the end of section 3.
By including the meson exchange potential,  
$V'_{(K+G)/N}+V_{m}$ (dot-dashed line) 
are also compared to $V'_{(K+G)/N+m}$ (solid line) in Fig.\ \ref{fig10}.
As seen in these figures, the equivalent local potentials 
given by solving the 
inverse scattering problem are almost additive in the  \NNcha\ channel.
This suggests that the nonlocality of the QCM potential
is rather weak in this channel,
 and that the potential 
may be simulated by the equivalent local potential, which
has the same on-shell behavior.

\begin{figure}
\includegraphics[width=7cm,clip]{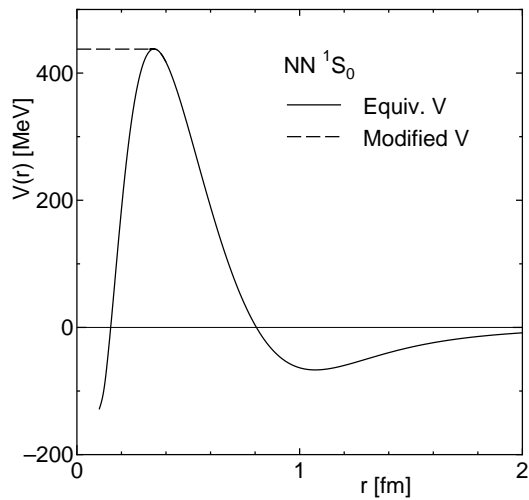}
\caption{Equivalent and modified local potential for the \NNcha\ channel 
in the OGEP quark model.}
\label{fig11}
\end{figure}

\begin{figure}
\includegraphics[width=7cm,clip]{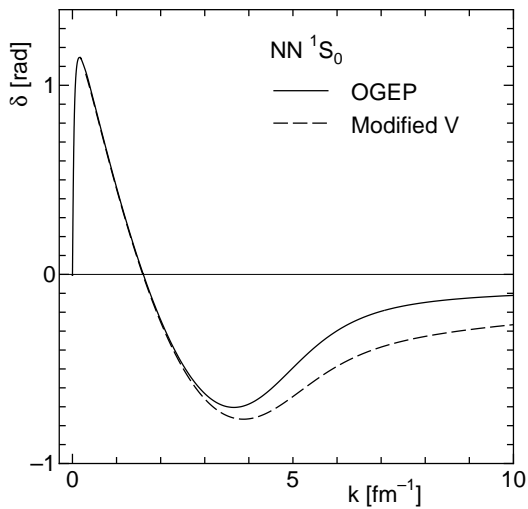}
\caption{Calculated phase shifts given by the equivalent and 
modified potentials 
for the \NNcha\ channel in the OGEP quark model.}
\label{fig12}
\end{figure}

Also, we investigate the effect of the short-range attractive pocket by 
performing a calculation without the pocket
in the NN system. In Fig. \ref{fig11}, we show 
 the modified 
local potential without the attractive pocket as well as the original
equivalent local potential with the attractive pocket. 
%
The phase shifts given by both of the two potentials are 
shown in Fig.\ \ref{fig12}.
%
From these figures, we understand that the modified potential gives the 
stronger repulsion at high energies. The difference between these phase 
shifts in the region from $k=$3 fm$^{-1}$ 
and higher produces the attractive 
pocket at short distances in the equivalent local potential.

\begin{figure}
\includegraphics[width=7cm,clip]{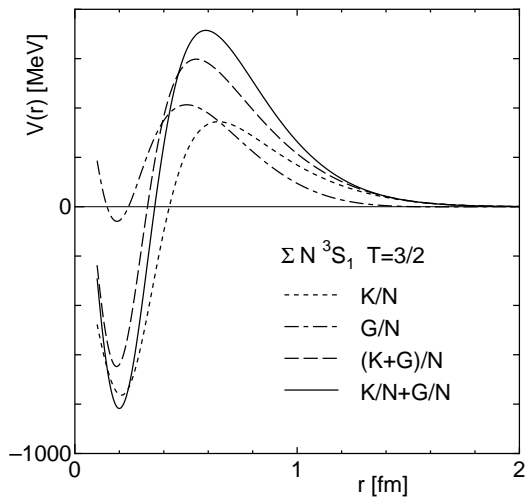}
\caption{Equivalent local potential for the \SNcha\ 
channel in the OGEP quark model. 
For further explanations see Fig.\ref{fig5} and \ref{fig9}.}
\label{fig13}
\end{figure}

\begin{figure}[t]
\includegraphics[width=7cm,clip]{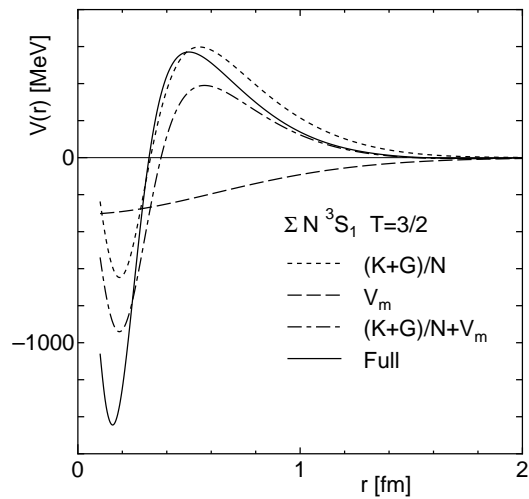}
\caption{Equivalent local potential for the \SNcha\ 
channel in the OGEP quark model. 
For further explanations see Fig.\ref{fig5} and \ref{fig10}.}
\label{fig14}
\end{figure}

In Figs. \ref{fig13} and \ref{fig14}, 
we again investigate the nonlocality of the QCM
potential in the \SNcha\ channel by checking 
whether the contributions from 
each term are additive or not.
%
As seen in the figures, the equivalent 
local potentials coming from each term shows the tendency to be additive. 
If we compare them with those for the \NNcha\ channel shown 
in Figs.\ \ref{fig9} 
and \ref{fig10}, however, we see that the additivity 
does not hold so well as in the   
\NNcha\ channel.
This suggests that the nonlocality of the QCM potential
is much higher in this channel. 
 
\section{Summary}

In this paper, we have investigated 
the short-range part of the potential given by the quark 
cluster model, especially the quark Pauli-blocking effects
and the nonlocal term coming from the quark potential. 
An energy-independent local potential 
can be reconstructed from a given
phase shift by solving the inverse scattering problem.
We have employed this method to derive
the energy-independent local potential which reproduces the 
same phase shifts as those obtained
from the QCM potential.

We have used two types of the quark cluster models.
One is called the OGEP quark model, where 
OGEP plays the dominant role to reproduce the mass difference 
between nucleon 
and $\Delta$, and also to produce the NN short-range repulsion. 
The other is the 
hybrid chiral (HC) quark model where the pseudoscalar- and 
scalar-meson-exchange 
potential between quarks including quark exchanges 
are taken into account together with OGEP.  
In the \NNcha\ channel, once the peak value is fitted,
there is almost no difference between these two models, but 
there appears a small difference in \SNcha channel. Since there is no 
essential difference between these two models, we have restricted ourselves 
to the OGEP quark model to investigate the details of the potential derived 
from the quark model.
 
We have found that such an equivalent local 
potential  has a strong repulsion at short distances in the \NNcha\
in both of the quark models. 
We have, however, also found there is an attractive pocket 
at very short distances.
It is considered that such a pocket appears because the 
nonlocal repulsion becomes weaker effectively at the very 
high energy region.
This repulsion-attractive-pocket 
structure becomes more manifest in the channel which has an 
almost forbidden state, \SNcha.  
There we have clarified the mechanism how the sharp increase of the phase 
shift occurs in connection with the norm kernel when there exists the almost 
forbidden state.  
From the shape of the equivalent local potentials, it is clearly seen that
the gluon exchange constructs the short-range repulsion in the \NNcha\
while the quark Pauli-blocking effect governs the feature of the 
scattering phase shifts in the \SNcha channel.
%

\begin{acknowledgements}
This work is supported in part by a Grant-in-Aid for Scientific Research
from JSPS (No.\ 11640258, 12640290)
\end{acknowledgements}


\appendix
\noindent
{\Large \bf Appendix} 
\section{Inverse Scattering Problem}
Here we briefly show how to derive an
equivalent local potential from a given phase shift: 
eqs.\ (\ref{eq22})-(\ref{eq25}).

We assume the S-matrix is given by the extended Eckert potential:
\begin{eqnarray}
S(k)&=&{g(k)+if(k) \over g(k)-if(k)} \label{eqA1}\\
\tan \delta(k) &=&f(k)/g(k)
\end{eqnarray}
with
\begin{eqnarray}
f(k)&=&k \sum_{m=0}^{n-1}f_{m}k^{2m}\\
g(k)&=& \sum_{m=0}^{n}g_{m}k^{2m}~.
\end{eqnarray}
The parameter $n$, \{$f_{m}$\}, \{$g_{m}$\} are determined
so that this S-matrix gives the given 
phase shift $\delta(k)$ up to about 15fm$^{-1}$.
Moreover, we use the condition,
$g_{n}=1$, $f(k_{0})=0$ at $\delta(k_{0})=0$ {\it etc}.,
and $f_{n}$ is determined to give the
Born term,
\begin{equation}
f_{n}=-{2\pi\mu\over \hbar^{2}}\bra V \ket.
\end{equation}

By using this form of the S-matrix, eq.\ (\ref{eq22})
becomes
\begin{eqnarray}
F(r)&=& \frac{1}{2\pi i} 
\int_{-\infty}^{+\infty} e^{ikr} {2 f(k) \over g(k)-if(k) } dk
\nonumber\\
&+&\sum_{{\rm all}~\kappa_j>0} c_j^2e^{-\kappa_j r},
\label{eqA2}
\end{eqnarray}
Because $g(k)-if(k)$ is the polynomial function of $k$ 
of up to the 2$n$-th order,
we have 2$n$ poles at $k=t_{i}$ ($i$=1 to 2$n$) 
in the integrand of the above equation in general. Thus, we have
\begin{eqnarray}
F(r)&=&  \sum_{{\rm all}~\Im t_{m}>0}  Res{2 f(t_{m}) 
\over g(t_{m})-if(t_{m})} e^{it_{m}r} 
\nonumber\\
&+&\sum_{{\rm all}~\kappa_j>0} c_j^2e^{-\kappa_j r},
\label{eqA3}
\end{eqnarray}

So the equation to solve, eq.\ (\ref{eq24}), becomes
a simple integral equation
with the known function, $F$, which can be solved numerically.
The potential in eq.\ (\ref{eq25}) is derived directly from the
kernel $K$.

The local potential which gives a given phase shift
is uniquely determined.
Since the systems we concern in this paper 
do not have a bound state,
the equivalent local potential can also be obtained
by fitting the phase shift directly.
We also use this fitting to check the above method.
The reconstructed potentials by these two methods are 
similar to each other with small numerical error,
which cannot be distinguished in the figures.

\section{A model for an almost forbidden state}
Here we explain a simplified model to investigate the mechanism 
of the role of an almost forbidden state. 
This model is simplified in a way that
the quark exchange does not occur among 
the  $ns$ ($n> 0$)
states.

First we introduce a projection 
operator $P$ which selects the $0s$ state.
\[
P= |0s\ket\bra 0s|.
\]
Then we consider the one-body scattering problem 
given by the following normalization 
$N$ and hamiltonian $H$.
\begin{eqnarray*}
&{}&H \chi = EN \chi \\
&{}&N = 1-\alpha P \\
&{}&H = T-\alpha (PT+TP)+ \alpha PTP, 
\end{eqnarray*}
where $T$ is the kinetic energy operator, $E$ is the energy eigenvalue and 
$\alpha$ is a parameter which is smaller than 1.
When $\alpha=1$, the state $0s$ is called the forbidden state, and if 
$\alpha$ is close to 1, the state $0s$ is called the almost forbidden state.
Introducing the projection operator $Q=1-P$, we rewrite the $N$ and $H$ 
as follows.
\begin{eqnarray*}
&{}&N =  Q+(1-\alpha)P \\
&{}&H = QTQ+(1-\alpha)(QPT+PTQ+PTP),
\end{eqnarray*}
This shows that the $0s$ state in the 
normalization $N$ is reduced by a factor of $(1-\alpha)$ and
the matrix elements of the hamiltonian $H$ 
between the $0s$ and $ns$ states are
also reduced by the factor of $(1-\alpha)$.
Now let us calculate the $H/N$. When $\alpha < 1$, 
 $1/\sqrt{N}$ is given by 
\[
\frac{1}{\sqrt{N}}=Q+\frac{1}{\sqrt{1-\alpha}}P.
\]
Then we obtain
\[
\frac{1}{\sqrt{N}}H\frac{1}{\sqrt{N}}=
QTQ+PTP+ \sqrt{1-\alpha}(PTQ+QTP).
\]
This tells that only the non-diagonal matrix elements between the 
$0s$ and $ns$ ($n \ne 0$) are reduced by a factor of $\sqrt{1-\alpha}$. Therefore 
the state which contains a large part of the $0s$ component, 
namely, the low-energy scattering state, 
feels effectively a repulsion when 
the factor $\sqrt{1-\alpha}$ is smaller than 1.     

\end{document}